\documentclass[12pt]{article}
\title {Bubble formation in $\varphi^4$ theory in the thin-wall 
limit and beyond}
\author{Hatem Widyan\thanks{E--mail : hatem@ducos.ernet.in},
A.Mukherjee\thanks{E--mail : am@ducos.ernet.in},
N.Panchapakesan\thanks{E--mail : panchu@ducos.ernet.in} and
R.P.Saxena\thanks{E--mail : rps@ducos.ernet.in}  \\
	{\em Department of Physics and Astrophysics,} \\
	{\em University of Delhi, Delhi-110 007, India} 
	}

\begin {document}
\maketitle
%
% This is the Abstract.
%
\begin {abstract}
 Scalar field theory with an asymmetric potential is studied at zero
temperature and high-temperature for $\varphi^4$ theory 
with both $\varphi$
and $\varphi^3$ symmetry breaking. The equations of motion are solved
numerically to obtain O(4) symmetric and O(3) cylindrical symmetric
bounce solutions. These solutions control the rates for tunneling
from the false vacuum to the true vacuum by bubble formation. The
range of validity of the thin-wall approximation (TWA) is
investigated. An analytical solution for the bounce is presented,
 which reproduces the action in the thin-wall as well as the 
thick-wall limits.
\end{abstract}

%
%This is the INTRODUCTION 
%
\begin {section}*{I. INTRODUCTION}
\par
The problem of decay of a metastable state via quantum tunneling 
has important 
applications in many branches of physics, from condensed matter 
to particle 
physics and cosmology. The tunneling is not a perturbative effect. 
In the semi-classical approximation, the decay rate per unit volume is 
given by an expression of the form
\begin{equation}
  \Gamma \quad = \quad A \quad e^{-S_E} , \label{eq:decay}
\end{equation}
where $S_E$ is the Euclidean action for the bounce: 
the classical solution of the equation of motion with appropriate boundary 
conditions. The bounce has turning points at the configurations at which 
the system enters and exits the potential barrier, and analytic 
continuation to Lorentzian time at the exit point gives us the 
configuration of the system at that point and its subsequent evolution.  
The solution of the 
equation of motion looks like a bubble in four dimensional Euclidean space 
with radius R and thickness proportional to the coefficient of the 
symmetry breaking term in the potential. When there are more than 
one solution 
satisfying the boundary conditions, the one with the lowest $S_E$ 
dominates Eq.~(\ref{eq:decay}). 
 The prefactor A comes from Gaussian functional integration over small 
fluctuations around the bounce. The zero-temperature formalism is
well-developed \cite{Langer,Coleman,Glasser}. In particular, it has been
proved rigorously that the least action is given by the bounce which
is O(4) invariant \cite{Glasser}.

 Linde \cite{Linde} extended the formalism to finite temperatures. He
suggested that at temperatures much smaller than the inverse radius of the
bubble at zero-temperature, the bounces are periodic in the Euclidean 
time $\tau$
direction and widely separated. Beyond this temperature they start 
merging into one another producing what is known as 
``wiggly cylinder'' solutions. As one keeps 
increasing the temperature these wiggles smoothly straighten out, and the 
solution goes into an O(3) invariant cylinder (independent of 
Euclidean time $\tau$) solution that dominates the thermal 
activation regime.

 The model field theory considered by most authors is $\varphi^4$
theory with a symmetry-breaking  term proportional to $\varphi$.
However, it was shown \cite{Dine} that the leading
temperature-dependent correction to the tree-level scalar potential
is proportional to $\varphi^3$. Thus it becomes of interest to obtain
bounce solutions to $\varphi^4$ theory with $\varphi^3$ symmetry-breaking. 
 An interesting special case is the so called thin-wall approximation
(TWA), when the bubble radius $R$ is much larger than the thickness of
the bubble wall. In this limit, there is an analytical formula for
$S_E$ in terms of the wall surface energy, and the details of the 
field theory are unimportant. However, it would be nice to also have an
analytical interpolating form for the solution itself.
 In the TWA limit, we expect the bounce
solutions in the two theories to be same, but for sufficiently large
coupling constants for the asymmetric terms we expect differences.
Also, it is not clear a priori what the limit of validity of the TWA is.

 Some of the above questions have been addressed by Adams
\cite{adams}. By suitably scaling and shifting the field $\varphi$ as
well as the potential $U(\varphi)$, he puts a general quatric
potential in the form
\begin{equation}
 U(\varphi) = {1 \over 4}\varphi^4 - \varphi^3 + {1 \over 2} 
\delta \varphi^2 , \label{adam:1}
\end{equation}
 where $ 0 \le \delta \le 2$. He then computes the $O(4)$ and $O(3)$
bounce action for various values of $\delta$. The thin-wall limit
corresponds to $\delta=2$, and the TWA result is recovered in that
case. This throws up the fresh question: how does the difference
between $\varphi$ and $\varphi^3$ symmetry-breaking get reflected in
the values of $\delta$ ? What is a thin wall in each of the two cases?

 In this paper we address the above issues. We obtain accurate
numerical solutions for the zero-temperature and high-temperature
bounces for $\varphi^4$ theory with both $\varphi$ and $\varphi^3$ 
symmetry-breaking. We compute
the actions in each case, and find that, for a modest value of the
asymmetric coupling $f(=0.25)$, the action given by TWA formula agrees
to within $9 \%$ with that obtained from the numerical
solution for $\varphi$ breaking. We also relate these values to the
$\delta$ parameter of Adams \cite{adams}.
However, the agreement is considerably better for
$\varphi^3$ breaking than $\varphi$ breaking. In the former theory, we
verify that as $f$ is reduced the error in the TWA formula goes to
zero. We propose another criterion for the goodness of TWA, in terms
of the temperature $T_\star$ at which the actions of the O(4) and O(3)
solutions become equal. A numerical investigation shows that the TWA holds
up to $f \sim 0.30$. Finally, we present an analytical solution
which satisfies the equation of motion with parameters fixed by
demanding stationary action. This reproduces TWA results very well
and, in the thick-wall limit, is in fairly good agreement with the
numerical results.

 In Sec. II we review bubble formation theory at zero and finite
temperature, and discuss the thin-wall approximation
(TWA). In Sec. III we present our numerical results for O(4) and
O(3) cases for $\varphi$ and $\varphi^3$ type symmetry breaking
as well as our investigation of the region of validity of TWA. In
Sec. IV the analytical solution is presented in the limits of a
thin wall as well as a thick wall. Finally, Sec. V contains 
some concluding remarks.

\end{section}
\begin {section}*{II. BUBBLE FORMATION}
%
%This is the BUBBLE FORMATION section.
%
\indent Let us consider a scalar field theory with a Lagrangian density
\begin{equation}
\mathcal{L}(\varphi)  =  {1 \over 2}({\partial_{\mu}\varphi})^2 - 
U(\varphi) ,
\end{equation}
 where the potential $U(\varphi)$ has two minima at 
$\varphi_-$ (false vacuum) and $\varphi_+$ (true vacuum). 

 In the semi-classical approximation the barrier tunneling
leads to the appearance of 
bubbles of a new phase with $\varphi = \varphi_+$ as classical solutions
in Euclidean space (i.e., imaginary time $\tau$). To calculate the 
probability 
of such a process in quantum field theory at zero temperature, one should 
first solve the Euclidean equation of motion : 
\begin{equation}
 \partial_{\mu}\partial_{\mu}\varphi = {dU(\varphi) \over d\varphi}
\label{equa:motion} ,
\end{equation}
with the boundary condition 
$\varphi \to \varphi_-$ as $ \vec x^2+\tau^2 \to \infty$ , 
where $\tau$ is the imaginary time. The probability of tunneling per unit 
time per unit volume is given by 
\begin{equation}
\Gamma  =   A \quad  e{^{-S_E[\varphi]}} \label{eq:S4},
\end{equation}
where $ S_E[\varphi]$ is the Euclidean action corresponding to 
the solution of Eq.~(\ref{equa:motion}) and given by the 
following expression :
\begin{equation}
S_E[\varphi] =  \int d^4{x} \left[ {1 \over 2} ({\partial\varphi \over 
\partial\tau})^2 + {1 \over 2} ( \nabla\varphi)^2 + U(\varphi) \right] . 
\end{equation}
\indent It is sufficient to restrict ourselves to the O(4) symmetric 
solution $ \varphi(\vec x^2+\tau^2)$ , since it is this solution that 
provides the minimum of the action $ S_E[\varphi]$ \cite{Glasser}.
 In this case Eq.~(\ref{equa:motion}) takes the simpler form 
\begin{equation}
{d^2\varphi \over d\rho^2} + {3 \over \rho} {d\varphi \over d\rho} 
= {dU(\varphi) \over d\varphi } \label{equt:S4}, 
\end{equation}
where $\rho=\sqrt{\vec x^2+\tau^2}$, with boundary conditions
\begin{equation} 
\varphi \to \varphi_-  \quad as \quad \rho \to \infty ,\quad 
{d\varphi \over d\rho }= 0 \quad at \quad \rho = 0 \label{cond:1} .
\end{equation}
 We denote the action of this solution by $S_4$.

\indent Now let us consider the finite temperature 
case. Following \cite{Linde}, in order 
to extend the above mentioned results to high temperature, 
$ T \not = 0 $, it is sufficient to remember that quantum statistics of 
bosons (fermions) at $ T \not = 0$ is equivalent to quantum field theory 
in the Euclidean space-time, periodic (anti-periodic) in the ``time'' 
direction with period $ \beta = T^{-1}$. One should use the $T$-dependent 
effective potential $U(\varphi,T)$ instead of the zero-temperature one 
$U(\varphi)=U(\varphi,0)$. Instead of looking for O(4)-symmetric solution 
of Eq.~(\ref{equa:motion}), one should look for O(3)-symmetric 
(with respect to spatial 
coordinates) solutions, periodic in the ``time'' direction with period 
$\beta = T^{-1}$. At sufficiently large temperature compared to 
the inverse 
of the bubble radius R at $T=0$, the solution is a cylinder whose spatial 
cross section is the O(3)-symmetric bubble of new radius $R(T)$. In this 
case, in the calculation of the action $S_E(\varphi)$ the integration 
over $\tau$ is reduced simply to multiplication by $T^{-1}$, i.e., 
$S_E[\varphi]= T^{-1} \> S_3[\varphi]$. Here $S_E[\varphi]$ is 
the four-dimensional action and $S_3[\varphi]$ is 
the three-dimensional action corresponding to the O(3)-symmetric bubble
and given by :
\begin{equation}
 S_3[\varphi]  =   \int d^3 r \left[ {1 \over 2} {(\nabla \varphi)^2 } 
+ U(\varphi,T) \right]  \label{num61:61} .
\end{equation}
\indent To calculate $S_3(\varphi)$ it is necessary to solve the equation
\begin{equation}
{d^2\varphi \over dr^2} + {2 \over r }{d\varphi \over dr}= 
{dU(\varphi,T) \over d\varphi } \label{equ:S3}
\end{equation}
with boundary conditions 
\begin{equation}
\varphi \to \varphi_- \quad as \quad r \to \infty , \quad
{d\varphi \over dr} = 0 \quad at \quad r=0 .
\end{equation}
where $r = \sqrt{\vec x^2}$. The complete expression for the 
probability of tunneling per unit time per unit volume in the 
high-temperature limit ($T >> R^{-1}$) is obtained in analogy to 
the one used in \cite{Coleman} and is given by:
\begin{equation}
\Gamma(T)  =  A(T) \> e^{-S_3[\varphi,T]/T} .
\end{equation}
 In the theory of bubble formation , the interesting quantity 
to calculate is the probability of decay between $\varphi=\varphi_-$ 
and $\varphi= \varphi_+$ 
which are the two minima of $U(\varphi)$. There is an interesting case
 (in the sense that the action can be calculated analytically) when 
$U(\varphi_+)-U(\varphi_-)= \varepsilon$ is much smaller than 
the height of the barrier. This is known as the thin-wall approximation
(TWA). At $T=0$, in the TWA limit, the action $S_4$ of the 
O(4)-symmetric bubble is equal to 
\begin{eqnarray}
S_4 & = &2 \pi^2 \int_0^\infty d\rho \> \rho^3 \left[ {1 \over 2}
({d\varphi \over d\rho})^2 + U(\varphi) \right]  \nonumber \\ 
 & = & -{1 \over 2} \varepsilon \pi^2 R^4 + 2 \pi^2 R^3 S_1 \label{e:S_4}.
\end{eqnarray}
 Here $S_1$ is the bubble wall surface energy (surface tension), given by
\begin{equation}
 S_1 = \int_0^\infty d\rho \left[({d\varphi \over d\rho})^2 +
U(\varphi)\right] , \label{a:S_1}
\end{equation}
and the integral should be calculated in the limit 
$\varepsilon \to 0$ .

The bubble radius $R$ is calculated by minimizing $S_4$
with respect to $R$ and this gives us
\begin{equation}
R= {{3 S_1} \over \varepsilon} ,
\end{equation}
whence it follows that 
\begin{equation}
S_4={{27 \pi^2 S_1^4} \over {2 \varepsilon^3}} . \label{ac:ac}
\end{equation}
The condition for the applicability of TWA is the following
\begin{equation}
{{3 S_1} \over \varepsilon} >> (d^2{U(\varphi_-)}/d\varphi^2)^{-1/2} ,
\end{equation}
where $(d^2{U(\varphi_-)}/d\varphi^2)^{-1/2}$ is simply the order 
of magnitude of the bubble wall thickness. The results presented above 
were obtained by Coleman \cite{Coleman}.

 These results can be easily extended to the case $ T >> R^{-1}$
\cite{Linde}. To this end it is sufficient to take into account that
\begin{eqnarray}
S_3 & = & 4 \pi \int_0^\infty dr \> r^2 \left[ {1 \over 2}
({d\varphi \over dr})^2 + U(\varphi,T) \right]  \nonumber \\ 
 & = & {-{4 \over 3}} \varepsilon \pi R(T)^3 + 4 \pi R(T)^2 S_1(T)
 \label{equation:S_3} ,
\end{eqnarray} 
where $S_1(T)$ is the bubble wall surface energy (surface tension) 
at finite temperature and is given by:
\begin{equation}
S_1(T) = \int_0^\infty dr \left[ ({d\varphi \over dr})^2 + U(\varphi,T) 
\right].
\end{equation}
 As before, the integral should be calculated in the limit 
$\varepsilon \to 0$.

The bubble radius $R(T)$ is calculated by minimizing 
$S_3$ with respect to $R(T)$ and this gives us 
\begin{equation}
R(T)= {{2 S_1(T)} \over \varepsilon} , 
\end{equation}
whence it follows that 
\begin{equation}
S_3={{16 \pi S_1^3(T)} \over {3 \varepsilon^2}} . \label{ac1:ac1}
\end{equation}
The condition for the applicability of TWA is the following
\begin{equation}
{{2 S_1(T)} \over \varepsilon} >> (d^2{U(\varphi_-,T)}/d\varphi^2)^{-1/2}, 
\end{equation}
where $(d^2{U(\varphi_-,T)}/d\varphi^2)^{-1/2}$ 
is the order of magnitude of the bubble wall thickness at high 
temperature. 

\end{section}
\begin {section}*{III. NUMERICAL RESULTS}
%
% This is the NUMERICAL RESULTS 
%
We start with the Euclidean action at $T=0$,
\begin{equation}
S_E[\varphi] = \int d^4{x} \left[ {1 \over 2} ({\partial\varphi 
\over \partial\tau})^2 + {1 \over 2} ( \nabla\varphi)^2 
+ U(\varphi) \right] . \label{num1:1}
\end{equation}
 If we have O(4) symmetry, Eq.~(\ref{num1:1}) reduces to 
\begin{equation}
S_4= 2 \pi^2 \int_0^\infty d\rho \> \rho^3 \left[ {1 \over 2 } 
({d\varphi \over d\rho })^2 + U(\varphi) \right] . \label{num2:2}
\end{equation}
 We compute the action for different values of the parameter in the 
symmetry-breaking term in the potential $U(\varphi)$.

First we consider the following potential :
\begin{equation}
U(\varphi) = { \lambda \over 2}(\varphi^2 - \mu^2)^2 - F \varphi ,
\label{num3:3} 
\end{equation}
where $\lambda$ is the coupling constant and $\mu$ is the mass of 
the scalar field. We perform the following rescaling in the
action~(\ref{num2:2})
\begin{equation}
\varphi \to \varphi / \mu ,\quad \rho \to \rho\sqrt{\lambda \mu^2} ,
\quad f = {F \over {\lambda \mu^3}} , \label{num4:4}
\end{equation}
to get 
\begin{equation}
S_4= {2 \pi^2 \over \lambda} \int d\rho \> \rho^3 \left[ {1 \over 2} 
({ d\varphi \over d\rho})^2 + {1 \over 2}(\varphi^2 -1)^2 -f 
\varphi \right] . 
\label{num5:5}
\end{equation}
 The only adjustable parameter in the Lagrangian is $f$, so by 
covering the whole range we should be covering all relevant cases.

The equation of motion is now
\begin{equation}
{d^2\varphi \over d\rho^2}+ {3 \over \rho} 
{d\varphi \over d\rho} = 2(\varphi^3-\varphi)-f , \label{num6:6}
\end{equation}
and the boundary conditions are the usual ones : 
\begin{equation}
\varphi \to \varphi_- \quad as \quad \rho \to \infty ,\quad {d\varphi 
\over d\rho}=0 \quad at \quad \rho=0 .
\end{equation}
By solving Eq.~(\ref{num6:6}) numerically for different 
values of $f$, substituting the solution in Eq.~(\ref{num5:5})
 and integrating, we obtain the action for each value of $f$.

 At high temperature, we look for the O(3) symmetric solution with
cylindrical symmetry. Then Eq.~(\ref{num61:61}) takes the form
\begin{equation}
S_3=4 \pi \int_o^\infty dr \> r^2 \left[{1 \over 2} ({d\varphi \over dr })
^2 +U(\varphi) \right] . \label{num7:7}
\end{equation}
 Using the same potential~(\ref{num3:3}) and the
rescaling~(\ref{num4:4}), we get
\begin{equation}
S_3 = {{4\pi \mu} \over \sqrt\lambda} \int_0^\infty dr \> r^2 \left
[ {1 \over 2} ({d\varphi \over dr})^2 + {1 \over 2} (\varphi^2-1)^2 -f 
\varphi \right]  . \label{num8:8}
\end{equation}
The equation of motion is then
\begin{equation}
{ d^2\varphi \over dr^2 } +{2 \over r} 
{ d\varphi \over dr } = 2(\varphi ^3 -\varphi) - f , \label{num9:9}
\end{equation}
and the boundary conditions are the usual ones : 
\begin{equation}
\varphi \to \varphi_- \quad as \quad r \to \infty ,\quad {d\varphi \over 
dr }=0 \quad at \quad r=0 . 
\end{equation}
 Again, we solve Eq.~(\ref{num9:9}) numerically for different
values of $f$, substitute the solution in Eq.~(\ref{num8:8}) and 
integrate to obtain the action for each $f$.

 Table I shows our numerical results at $T=0$ and 
high temperature for different values of the asymmetry parameter $f$.
\begin{center}
Table I
\end{center}
\begin{center}
\begin{tabular}
{|p{1.2cm}|p{3.2cm}|p{3.2cm}| p{3.2cm} | p{3.2cm}|}
\hline
 f & $S_4$ (Numerical) & $S_3$ (Numerical) & $S_4$(Analytical) & 
$S_3$(Analytical) \\
\hline 
 0.25 & 3064.70 & 143.80 & 3368.80 & 158.90 \\ 
\hline 
 0.55 & 171.30 & 18.70 & --- & --- \\
\hline
0.75 & 13.20 & 1.60 & --- & --- \\ 
\hline 
\end{tabular}
\end{center}
\vskip 1 cm
 As we discussed in section II, for small values of $f$ we can use the
TWA formula for computing the action. From Eq.~(\ref{a:S_1})
\begin{eqnarray}
S_1 & = & \int_0^\infty dr \left[ ({d\varphi \over dr})^2 + {1 \over 2} 
(\varphi^2 -1 )^2 \right] \nonumber \\
& = & - \int_{-1}^1 d\varphi \> \sqrt{(\varphi^2-1)^2} \nonumber \\
 & = & {4 \over 3}  \label{10:10}
\end{eqnarray}
 The radius is given by 
\begin{equation}
R = {{3 S_1} \over \varepsilon }, 
\end{equation}
where $ \varepsilon = 2f$ (see \cite{Coleman}). 
For $f=0.25$, we have $ R=8$ and the value of the action is (see
Eq.~(\ref{ac:ac}))
\begin{equation}
S_4 = 3368.80 .  \label{num11:11}
\end{equation}
Comparing this analytic value with the numerical value for $f=0.25$, 
 we get an error equal $9\%$.

At high temperature, again $S_1= 4/3$. The value of 
$R(T)= 16/3$ and the action is (see
Eq.~(\ref{ac1:ac1}))
\begin{equation}
S_3(T)= 158.90 . \label{num12:12}
\end{equation}
 Comparing this analytic value with the numerical value for 
$f=0.25$, we get
an error equal $9 \%$. Thus even for $f$ as small as $0.25$
the TWA formula for the action does not give very accurate
results. Obviously, there is no point in comparing numerical results
obtained for higher values of $f$ with the TWA formula.

Now we repeat the above calculations for the following potential 
\begin{equation}
U(\varphi)= {\lambda \over 2}( \varphi^2 - \mu^2)^2 -F \varphi^3 . 
\label{num13:13}
\end{equation}
We scale $\varphi$ and $\rho$ as before, with the dimensionless
coupling $f$ defined by $f =F/(\lambda \mu)$, solve the equation of
motion and calculate the action. Table II shows our 
numerical results for different values of $f$.
\begin{center}
Table II
\end{center}
\begin{center}
\begin{tabular}
{|p{1.2cm}|p{3.2cm}|p{3.2cm}| p{3.2cm}| p{3.2cm}|}
\hline
 
f & $S_4$ (Numerical) & $S_3$ (Numerical) & $S_4$(Analytical) & 
$S_3$(Analytical) \\
\hline 
 0.25 & 3446.10 & 161.80 & 3368.80 & 158.90 \\ 
\hline 
 0.55 & 349.50 & 35.40 & --- & --- \\
\hline
0.75 & 139.20 & 19.90  & --- & --- \\ 
\hline 
\end{tabular}
\end{center}
\vskip 1 cm

We now check the TWA formula for $f=0.25$. From table II we see that
the error in $S_4$ is only $2.3 \%$, while the error in $S_3$ is
$1 \%$. 
 Thus we see that for $f=0.25$, the TWA holds much better for symmetry
breaking with a $\varphi^3$ term than with a $\varphi$ term.
  
 Notice that the TWA formula is independent of the symmetry-breaking
term in the 
potential. Moreover, the action for large values of $f$ for the 
$\varphi^3$ symmetry-breaking term is larger than that for the 
$\varphi$ one for both the O(4) and the O(3) cases.

 To test our numerical method (we have used Hamming's modified
predictor-corrector method for solving the equation of motion), we 
have calculated the action for small
values of the symmetry breaking parameter $f$ in the
potential~(\ref{num13:13}) and compared it with the TWA
formula. In Fig. (1), we plot the percentage error in the TWA formula
as a function of $f$. The dots represent our results while the solid
line shows a fit to the data. We see that the error decreases for
small $f$, as expected, and approaches zero as $f \rightarrow 0$. 

 Table III shows the relationship between the values of
$\delta$ in Eq.~(\ref{adam:1}) and $f$ for $\varphi$ and $\varphi^3$
perturbation.
\begin{center}
Table III
\end{center}
\begin{center}
\begin{tabular}
{|l|lr|} \hline
$f$  & $\quad \quad ~ ~ ~\delta$ & \\ \cline{2-3}
& $\varphi$ case & $\varphi^3$ case \\ \hline
 0.0 & 2.0 & 2.0 \\ \hline
 0.25 & 1.85 & 1.85 \\ \hline
 0.55 & 1.50 & 1.62 \\ \hline
 0.75 & 0.64 & 1.46 \\ \hline
 0.77 & 0.0  & 1.44  \\ \hline
 0.80 & ---  & 1.41 \\ \hline
 1.0  & ---  & 1.25 \\ \hline
 5.0  & ---  & 0.15 \\ \hline
 10.0 & ---  & 0.03 \\ \hline
\end{tabular}
\end{center}

 As we can see from the table, as $f$ goes from $0$ to $0.77$,
$\delta$ goes from $2$ to $0$ for the $\varphi$ case (for $f > 0.77$ 
the maximum disappears) and from $2$ to
$1.44$ for the $\varphi^3$ case. Also we can see that $\delta$ goes 
to $0$ for the $\varphi^3$ case only at large values of $f$ (here the
maximum disappears only asymptotically).

 As already mentioned, at zero temperature the O(4) symmetric solution
has the lowest value of $S_E$, i.e., $S_E=S_4$. At high temperature,
we have $S_E=S_3/T$. At intermediate temperatures other solutions
exist. In the TWA, however, it has been shown \cite{Garriga} that all
other solutions have higher Euclidean action. This corresponds to a
first order phase transition from quantum tunneling at low temperature
to thermal hopping at high temperatures. The transition temperature
$T _\star$ is given by equating $S_4$ with $S_3/T$, i.e., 
\begin{equation}
T_\star = {S_3 \over S_4} \label{num15:15}
\end{equation}
If the surface tension $S_1$ is temperature independent, we have 
\begin{equation}
S_4  =  {{27 \pi^2 S_1^4} \over {2 \varepsilon^3}} \label{num16:16}
\end{equation}
\begin{equation}
S_3 =  {{16 \pi S_1^3} \over {3 \varepsilon^2}} \label{num17:17}
\end{equation}
Dividing  Eq.~(\ref{num16:16}) by Eq.~(\ref{num17:17}) and putting
$\varepsilon=2f$ (see \cite{Coleman}) we get
\begin{equation}
T_\star = C * f
\end{equation}
where
\begin{equation}
C = {64 \over {81 \pi S_1}}
\end{equation}
Thus we see that, in the TWA, $T_\star$ increases linearly with $f$. We
test this by computing $S_3/S_4$ from our numerical solutions at
different values of $f$. Fig. 2 shows our results for the potential of
Eq.~(\ref{num13:13}). We see that, for $f \leq 0.3$, there is very
good agreement with the predicted linear dependence. This also
confirms that, in the domain of validity of the TWA, the surface
tension $S_1(T)$ is independent of $T$. Beyond $f \sim 0.3$ in our
dimensionless units, there is a systematic deviation from
linearity. Thus we can say that, for values of $f$ larger than this,
the wall thickness becomes important.

\end{section}
\begin{section}*{IV. ANALYTIC SOLUTION IN TWA LIMIT}
%
% This is the analytic part.
%
\def\t{e^{{{(\rho^2 -R^2)}/  \Lambda^2}} + 1}
\def\g{\gamma}
\def\d{\delta}
\def\l{\Lambda}
 We use the following potential (see \cite{adams}) to calculate
the action analytically in two extreme limits: the thin-wall and
thick-wall limits.
\begin{equation}
U(\varphi) = {1 \over 4}\varphi^4 - \varphi^3 + 
{1 \over 2} \delta \varphi^2 .
\end{equation}
\noindent
\underline{\it{ Thin-wall limit : $\d \to 2$}}

We find that an analytic solution for the bounce of the form of a Fermi
function:
\begin{equation}
\varphi =  {\g \over \t} \label{twa1:1},
\end{equation}
where  $\rho=\sqrt{\vec x^2+\tau^2}$, $R$ is the radius of the bubble
and $\l$ its width, acts like a bounce 
in the TWA and leads to the correct value for the action $S_4$.   

Here the false minimum of the potential is at $\varphi=0$ and the true
minimum  lies between 2 (for $\d=2$) and 3 (for $\d=0$). The parameter
$\g$ is approximately equal to true minimum in the TWA.
The bounce has values $\varphi = \g$ at $\rho=0$ 
and $0$ at $\rho \to \infty$. The boundary conditions~(\ref{cond:1})
are satisfied by Eq.~(\ref{twa1:1}).

 We discuss the zero-temperature action $S_4$. 
To evaluate $\g$, $R$, and $\l$, we substitute
the ansatz (\ref{twa1:1}) in Eq.~(\ref{equt:S4}) :
\begin{equation} 
{d^2\varphi \over d\rho^2}+ {3 \over \rho}{d\varphi \over d\rho} = 
\varphi^3 - 3 \varphi^2 + \d \varphi . \label{twa2:2}
\end{equation}
Then the left-hand side (L.H.S.) and the right-hand side (R.H.S.) are
respectively
\begin{eqnarray} 
L.H.S.= {{\g 8 \rho^2 / \l^4} \over (\t)^3} & + &{{\g (-12 \rho^2 / \l^4 
+ 8 / \l^2)} \over (\t)^2 }  \nonumber \\ [0.3cm]
 & + & {{\g (4 \rho^2 / \l^4 - 8 / \l^2)} \over \t }
\label{twa3:3} . 
\end{eqnarray}
\begin{eqnarray}
R.H.S.= {{\g \d} \over \t}
 & - & {{ 3 \g^2} \over (\t)^2 } \nonumber \\ [0.3cm]
 & + & {{\g^3} \over (\t)^3}
\label{twa4:4} .
\end{eqnarray}
 In the TWA, the solution is constant except in a narrow region near the 
wall at $\rho=R$. So, we replace in Eq.~(\ref{twa3:3}) 
\begin{eqnarray}
\mathrm{
8 \rho^2 / \l^4 \quad by \quad {{ 8 R^2} \over \l^4} ( 1- a \l^2 /R^2) 
\quad in \quad the \quad { 1 \over (\t)^3} \quad term }
\label{twa5:5} , \\ [0.3cm]
\mathrm{
8 / \l^2 -12 \rho^2 / \l^4 \quad by \quad {-}{{ 12 R^2} \over \l^4}
(1- b \l^2 / R^2 ) \quad in \quad the \quad { 1 \over (\t)^2} \quad 
\quad term }, \label{twa6:6} \\ [0.3cm]
\mathrm{
4 \rho^2 / \l^4- 8 / \l^2 \quad by \quad {{4 R^2} \over \l^4}
(1- d  \l^2 / R^2) \quad in \quad the \quad { 1 \over \t} \quad term
}
\label{twa7:7} ,
\end{eqnarray}
where $a$, $b$ and $d$ are parameters to be determined later.

Comparing Eq.~(\ref{twa3:3}) with Eq.~(\ref{twa4:4}) in the
range
$ R^2(1- \l^2 / R^2)=R^2-\l^2 < \rho^2 < R^2 + \l^2 =R^2(1+ \l^2 / R^2)$ 
where $\rho^2 \simeq R^2$  as $\l^2 / R^2 << 1$ , we have :
\begin{eqnarray}
{\g^2 \over 8} = {{ R^2} \over \l^4} (1- a \l^2 / R^2) , 
\nonumber \\ [0.3cm]
{\g \over 4} = {{R^2} \over \l^4}( 1-b \l^2 / R^2) ,
\label{twa10:10} \\ [0.3cm]
{ \d \over 4} = {{R^2 \over \l^4} ( 1- d \l^2 / R^2)} ,
\nonumber 
\end{eqnarray}
We can now evaluate the zero-temperature action $S_4$ : 
\begin{equation}
S_4= 2 \pi^2 \> \int_0^\infty d\rho \> \rho^3 \left[{1 \over 2} 
({d\varphi \over d\rho})^2 + U(\varphi) \right] .
\label{twa12:12} 
\end{equation}
Substituting Eq.~(\ref{twa1:1}) in Eq.~(\ref{twa12:12}) and
integrating we get
\begin{eqnarray}
S_4 & = & 2 \pi^2 \g^2 R^4 \Bigg[ { 1 \over {6 \l^2}}(1+({\pi^2 \over 3} -
2) { \l^4 \over R^4}) + {\d \over 8} (1- 2 \l^2 / R^2 
 +  { \pi^2 \over 3}{ \l^4 \over R^4})  \nonumber \\
& & + {\g \over 4} (1- 3 \l^2 / R^2 + ({\pi^2 \over 3} + 1) \l^4/R^4) 
\nonumber \\
& & + { \g^2  \over 16} (1- {{11 \l^2} \over {3 R^2}}+ ({\pi^2 \over 
3} + 2) \l^4/R^4) \Bigg].  \label{twa13:13}
\end{eqnarray}
We now determine the parameters $a$, $b$, and $d$ by demanding 
${dS_4 / dR^2}={dS_4 / d\l^2}={dS_4 / d\g}=0$.
 Differentiating Eq.~(\ref{twa13:13}) and using
Eq.~(\ref{twa10:10}), we find that to 
leading order in $\l^2/R^2$,
\begin{eqnarray}
4b- 2a -2d +1 =0, \nonumber \\ [0.3cm]
3b -2a - d =0 ,  \\ [0.3cm] \label{twa14:14}
3b- 11a /6 -d =0, \nonumber
\end{eqnarray}
which leads to $a=0$, $b=1/2$ and $d=3/2$. Using Eq.~(\ref{twa10:10}),
we can rewrite Eq.~(\ref{twa13:13}) as :
\begin{eqnarray} 
S_4 = {2 \pi^2}{ 8 R^6 \over \l^6}\Bigg[ (1/3-d/2-a/2+b) & + & 
{\l^2 \over R^2} (d - 3b + 11 a /6) \Bigg] , \label{twa15:15}
\end{eqnarray}
where the coefficient $\l^4 \over R^4$ evaluated by the usual methods 
of statistical mechanics for the Fermi function vanishes. This gives
\begin{equation}
S_4 = {4 \pi^2 \over 3} { R^6 \over \l^6} +O({\l^6 \over R^6}) . 
\label{twa16:16}
\end{equation}
The quantities $\g$, $R$ and $\l$ are determined from
Eq.~(\ref{twa10:10}) using the values of $a$, $b$, and $d$. So we have 
\begin{equation}
\g^2 - 2 \g {{d-a \over {d-b}}} + 2 \d {{b-a \over d-b}} = 0 ,
\label{twa25:25}\\ [0.3cm]
\end{equation}
which gives
\begin{equation}
\l^2 = {8(b-a) \over {\g^2-2\g}}={{4 (d-b) \over {\g -\d}}} , 
\label{twa17:17}
\end{equation}
with $\g$ given by Eq.~(\ref{twa25:25}). We have then, for
$\d=1.9$, $\g=2.1$, which implies that $ R^2/\l^2={\g \over {\g-\d}}+
b = 11$. Thus we have
\begin{equation}
 S_4 = { 4 \pi^2 \over 3} (11)^3 , \label{twa18:18}
\end{equation}
while the action from the TWA formula is ( see \cite{adams}) 
$S_{TW} = { 4 \pi^2 \over 3} (10)^3$ for $\d=1.9$.  The departure from
TWA, ${S_4 / S_{TW}}=1.33$, is in agreement with 
Ref.~\cite{adams}. The expressions seem certainly valid 
for values of $\d$ in the range 2.0 to 1.8 .
\vskip 1.2 cm
\noindent
\underline{\it{ Thick-wall limit: $\d  \to 0$}}
 
 The form of the bounce in Eq.~(\ref{twa1:1}) suggests that the thick
wall limit, which would correspond to small values of $R^2/\l^2$, would
be obtained by approximating the Fermi function by the 
Maxwell-Boltzmann function, which leads to a Gaussian:
\begin{equation}
 \varphi = \g e^{-\rho^2/\l^2} . \label{twa19:19}
\end{equation}
 The action for this form of bounce is found to be
\begin{equation}
 S_4 = { {\pi^2 \g^2 \l^4}} \Bigg[ {1 \over {2\l^2}} 
+ { \d \over
8} - {\g \over 9} + {\g^2 \over 64}\Bigg]. \label{twa20:20}
\end{equation}
Eqs.~(\ref{twa10:10}) then reduce to 
\begin{equation}
{\g^2 \over 8} = -{a \over \l^2} ~, \quad 
{\g \over 4} = -{ b \over \l^2} ~, \quad 
{ \d \over 4} = - {d \over\l^2} ~.
\end{equation}
Note that in this case $\g \ll 1$, so $\g^2$ is negligible.

The values of $b$ and $d$ are again obtained by demanding
$dS_4/d\l^2=dS_4/d\g = 0$. This gives $b=-9/8$, $d=-1/2$, giving
\begin{equation}
\l^2 = {2 \over \d} ~, \quad \g={9 \over 4} \d .
\end{equation}
This yields the action
\begin{equation}
 S_4 = {{4 \pi^2} \over 3} (1.9) \d + O({R^2 \over \l^2}) . 
\label{dwt1:1}
\end{equation}
 The ratio of the action to the TWA value is 
\begin{equation}
 R_4 = { S_4 \over S_{TW}} = 1.9 \d (2-\d)^3 .
\end{equation}
For $\d=0.1$, $R_4=1.31$, which agrees with Adams' result.

 Thus, the form of the bounce given by Eq.~(\ref{twa10:10}) seems valid
over the whole range of $\d$ ( from 0 to 2 ), and in the two extreme
limits is amenable to analytic calculations.

 The numerical plots of the function $\varphi$ in the two extreme
limits are shown in Figures 3 and 4. In Fig. 3 we compare our numerical
result with the analytic one for $\d=1.8$. From the figure we see
that the Fermi function agrees very well with our numerical
results. Similarly, Fig. 4 shows the Gaussian function compared
with our numerical results for $\d=0.2$. We find that $\g=0.55$ 
numerically and $0.45$ analytically. This discrepancy in the values of
$\g$ is expected, because the terms neglected in Eq.~(\ref{dwt1:1}) are of
order $({R^2 \over \l^2})$.

 The analysis given above can be carried out for the O(3) symmetric
bounce and we find that this gives equally good results. We anticipate
that this approach will be of great use is discussing the extension to
arbitrary temperature. This will be discussed in a future publication.

\end{section}
\begin{section}*{V. CONCLUSIONS}
%
% This is the conclusion.
%
 We have obtained accurate numerical solutions for the
zero-temperature and high-temperature bounces for
both $\varphi$ and $\varphi^3$ symmetry-breaking. We compute
the actions in each case and find that, for a modest value of the
asymmetric coupling $f=0.25$, the action given by the TWA formula agrees
to within $9 \%$ with that obtained from the numerical
solution for $\varphi$ breaking while for $\varphi^3$ the agreement is
within $2.3 \%$. Hence, the agreement is
considerably better for $\varphi^3$ breaking than $\varphi$ breaking.
At high temperatures, the conclusion is qualitatively similar.

 We have checked our numerical method by comparing the action
obtained numerically with the one obtained from the TWA formula. 
Very good agreement is obtained as we go to small values of $f$. We also 
verify that as $f$ is reduced the error in the TWA formula goes to
zero. We propose an alternative criterion for the goodness of TWA, 
in terms of the relation between $f$ and 
the temperature $T_\star$ at which the actions of the O(4) and O(3)
solutions become equal. A numerical investigation shows that TWA holds
up to $f \sim 0.30$. Finally, we present an analytical solution
which satisfies the equation of motion in an approximate sense in two
limiting cases. The
first of these reproduces the leading corrections to the 
TWA results very well. The second is applicable for the
opposite case of a very thick wall. This gives us insights into the
nature of the bounce solutions for various values of $\delta$ going
from thin to thick walls. 

Our work overlaps to some extent with that of Adams
\cite{adams}. Like us, he had computed the bounce action for both
$O(4)$ and $O(3)$ symmetry. Using the parametrization of 
Eq.~(\ref{adam:1}),
he was able to map the entire range of wall thickness into the
interval $ 0 \le \delta \le 2$. However, while the form
~(\ref{adam:1}) is computationally very convenient, it obscures the
fact that the free parameter actually measures the departure from pure
$\varphi^4$ theory, and does not directly allow a comparison between
$\varphi$ and $\varphi^3$ symmetry-breaking. Our parametrization
allows such a comparison. Of course, both theories can be mapped onto
~(\ref{adam:1}), but the $\delta-f$ relationship is different, as
illustrated in Table III. Moreover, Adams does not attempt to obtain 
approximate
analytical solutions, which is done here for the first time. Also, he
does not lay any emphasis on the $T_\star-f$ relationship and indeed does
not compute $T_\star$.

Much of the work on inflationary 
models relies on the zero-temperature potential, so our results could be
relevant for inflation \cite{Linde}. They may also have some bearing
on the formation of topological defects in a first order phase
transition where authors consider zero-temperature potentials, see for
example \cite{Digal}.

 So far, we have discussed the action only at zero temperature and high  
temperature. To obtain the bounce solution at intermediate
temperatures, we have to solve a partial differential equation with
periodic boundary conditions in the $\tau$ direction. We have done so
using a multigrid method as well as by extending the analytic bounce
calculations. This work will be  presented in a future publication. 

\end{section}
\begin{section}*{ACKNOWLEDGEMENTS}
% Ack. 
%
This work is part of a project (No. SP/S2/K--06/91) funded by the
Department of Science and Technology, Government of India. H.W. 
thanks the University Grants Commission, New Delhi, for a fellowship.  
 We thank a referee for drawing our attention to Ref.\cite{adams}.

\end{section}
\bibliography{plain}

\newpage
\begin{section}*{Figure Caption}
FIG. 1. Error in the TWA formula as a function of $f$. 
The squares represent our results while the solid 
line shows a fit to the data.

FIG. 2.  Deviation of $T_\star$ from the TWA limit.
The straight line represents the TWA limit  
while the squares are our numerical results.

FIG. 3. $\varphi$ as a function of $\rho$. The solid line is the
Fermi function while the dashed line is the numerical result.

FIG. 4. $\varphi$ as a function of $\rho$. The solid line is the
Gaussian function while the dashed line is the numerical result.

\end{section}

\newpage
\begin{figure}[ht]
\vskip 15truecm

\includegraphics{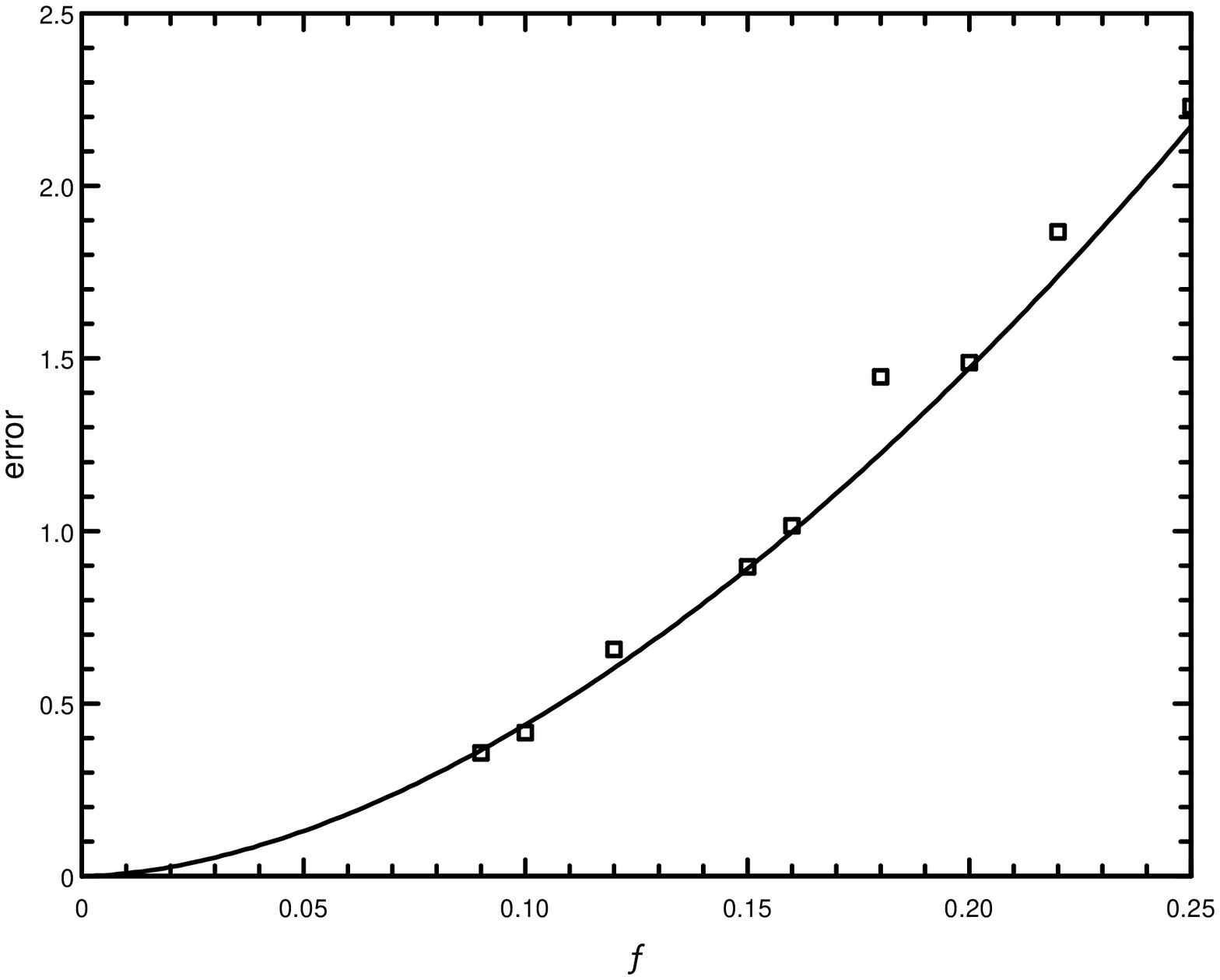} 

\caption{}
\end{figure}
\newpage
\begin{figure}[ht]
\vskip 15truecm
 \includegraphics{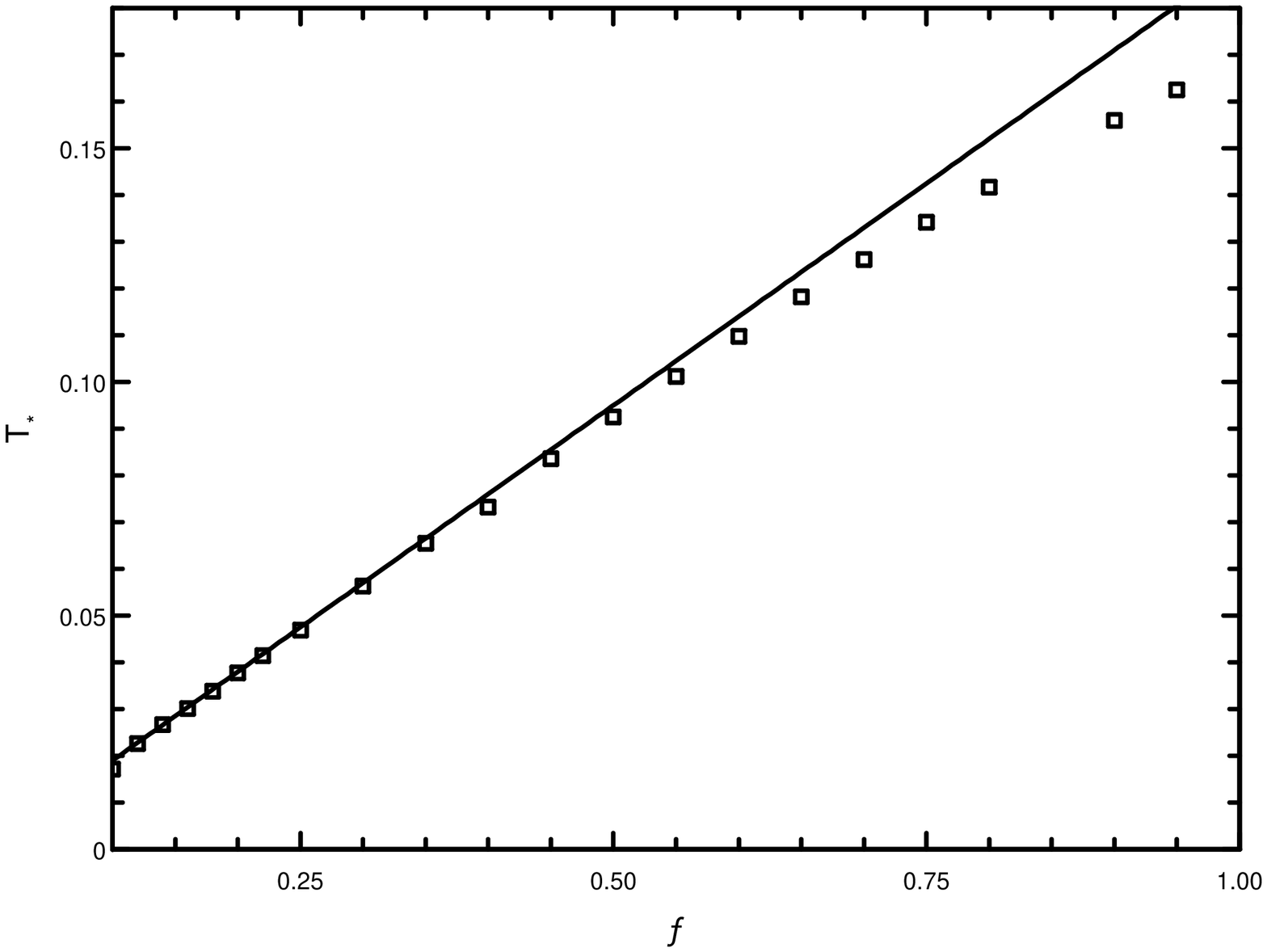} 
\caption{}
\end{figure}
\newpage
\begin{figure}[ht]
\vskip 15truecm
 \includegraphics{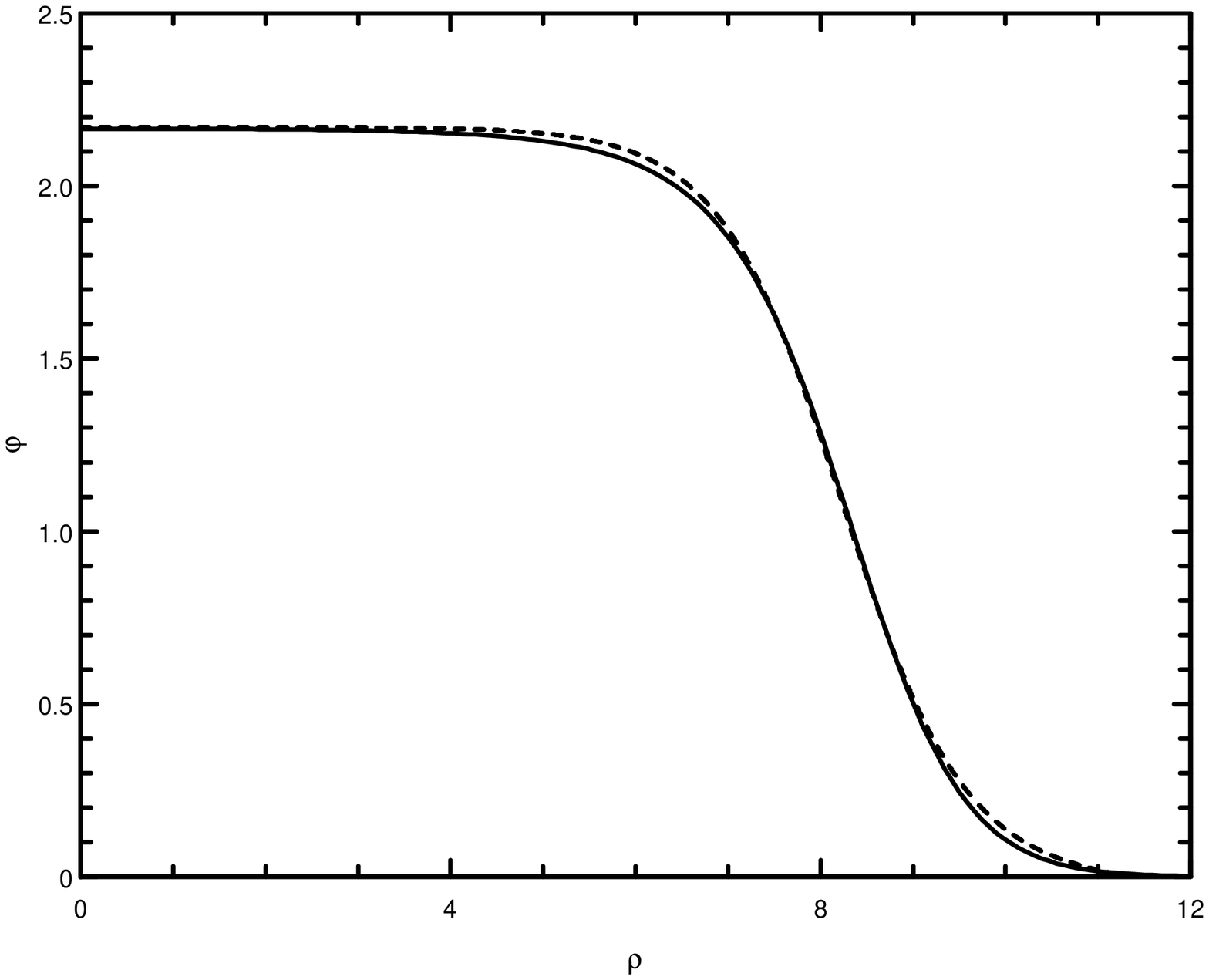} 
\caption{}

\end{figure}
\newpage
\begin{figure}[ht]
\vskip 15truecm
 \includegraphics{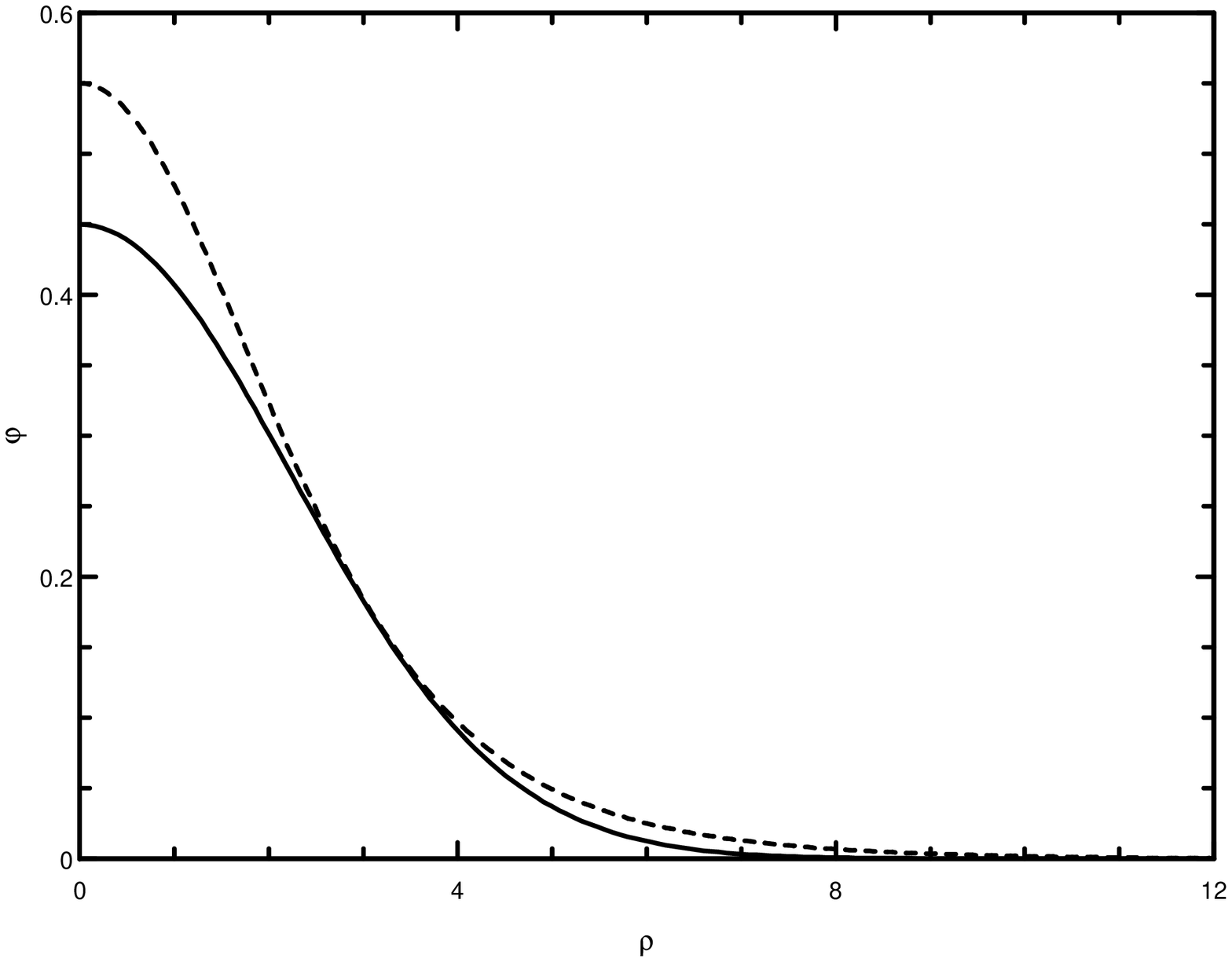} 
\caption{}

\end{figure}
\end{document}